\documentclass[a4paper,authoryear]{cas-dc}
\usepackage{graphicx}
\usepackage{multirow}
\usepackage{booktabs}
\usepackage{enumitem}
\usepackage{color}
\usepackage{subfig}
\usepackage{float}
\usepackage{wrapfig}
\usepackage{tabularx}
\usepackage{mathbbol}
\usepackage{amsmath}
\usepackage{amssymb}
\usepackage{colortbl}
\usepackage{stfloats}
\usepackage{float}
\usepackage{color}
\usepackage[ruled]{algorithm2e}
\usepackage[shortcuts]{extdash}

\usepackage{newfloat}
\usepackage{listings}
\usepackage{natbib}

\def\tsc#1{\csdef{#1}{\textsc{\lowercase{#1}}\xspace}}
\tsc{WGM}
\tsc{QE}
\tsc{EP}
\tsc{PMS}
\tsc{BEC}
\tsc{DE}

\pdfoutput=1

\begin{document}
\let\printorcid\relax
\let\WriteBookmarks\relax
\def\floatpagepagefraction{1}
\def\textpagefraction{.001}
\shorttitle{Preprint submitted to Elsevier}
\shortauthors{Meng et~al.}

\title [mode = title]{Continuous K-space Recovery Network with Image Guidance for Fast MRI Reconstruction}                      

\author[1,2]{Yucong Meng}[style=chinese]
\ead{ycmeng21@m.fudan.edu.cn}
\author[2,3]{Zhiwei Yang}[style=chinese]
\ead{zwyang21@m.fudan.edu.cn}
\author[1,2]{Minghong Duan}[style=chinese]
\ead{mhduan22@m.fudan.edu.cn}
\author[1,2]{Zhijian Song}[style=chinese]\cormark[1]
\ead{zjsong@fudan.edu.cn}
\author[1,2]{Yonghong Shi}[style=chinese]\cormark[1]
\ead{yonghong.shi@fudan.edu.cn}
\affiliation[1]{organization={Digital Medical Research Center, School of Basic Medical Science,Fudan University}, 
                city={Shanghai},
                postcode={200032}, 
                country={China}}
\affiliation[2]{organization={The Shanghai Key Laboratory of Medical Image Computing and Computer Assisted Intervention}, 
                city={Shanghai},
                postcode={200032}, 
                country={China}}
\affiliation[3]{organization={Academy of Engineering and Technology, Fudan University}, 
                city={Shanghai},
                postcode={200433}, 
                country={China}}

\cortext[cor1]{Corresponding author}
\begin{abstract}
Magnetic resonance imaging (MRI) is a crucial tool for clinical diagnosis while facing the challenge of long scanning time. To reduce the acquisition time, fast MRI reconstruction aims to restore high-quality MRI images from the undersampled k-space. Existing methods typically train deep learning models to map the undersampled data to artifact-free MRI images. However, these studies often overlook the unique properties of k-space and directly apply general networks designed for image processing to k-space recovery, leaving the precise learning of k-space largely underexplored. In this work, we propose a continuous k-space recovery network from a new perspective of implicit neural representation with image domain guidance, which boosts the performance of MRI reconstruction. Specifically, (1) an implicit neural representation based encoder-decoder structure is customized to continuously query unsampled k-values. (2) an image guidance module is designed to mine the image-domain information from the low-quality MRI images to further guide the k-space recovery. (3) a multi-stage training strategy is proposed to recover dense k-space progressively. Extensive experiments conducted on CC359, fastMRI, IXI, and SKM-TEA datasets demonstrate the effectiveness of our method and its superiority over other competitors. Implementation code will be available in \href{https://github.com/XiaoMengLiLiLi/IGKR}{https://github.com/XiaoMengLiLiLi/IGKR}.
\end{abstract}

\maketitle
\section{Introduction}
Magnetic resonance imaging (MRI) is widely used for clinical diagnosis because of its advantages of no radiation, high resolution, and satisfactory soft-tissue contrast. However, the acquisition of dense k-space for accurate MRI images inherently leads to long time due to both physiological and hardware limitations~\citep{r1,r2}. The extended scan time causes patient discomfort and reduces accessibility to MRI. To speed up this process, accelerated MRI aims to reconstruct MRI images from undersampled k-space~\citep{r3,r4,r5}. However, the aliasing artifacts caused by such insufficient sampling often affect the clinical diagnosis. Therefore, it becomes a significant challenge how to reduce the amount of k-space acquisition while maintaining or even improving MRI image quality.
\begin{figure}[!t]
\centerline{\includegraphics[width=1.0\linewidth]{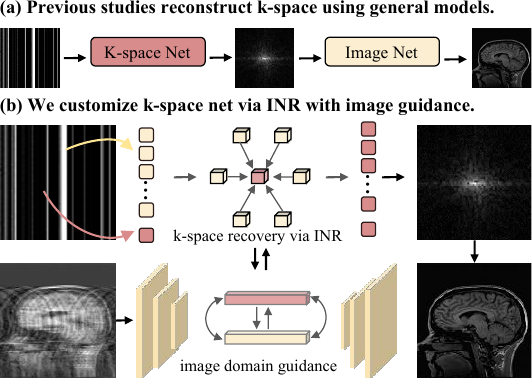}}
\caption{Our main idea. (a) Previous studies typically employ general networks to reconstruct the undersampled data both in the k-space and image domain. (b) We customize a continuous k-space recovery network from a new perspective of implicit neural representation with image domain guidance, thereby enhancing the performance of MRI reconstruction.}
\vspace{-0.5 cm}
\label{Fig.1}
\end{figure}

In recent years, deep learning has emerged as a powerful tool to tackle this problem. These methods usually learn the mapping between low-quality MRI images and their corresponding high-quality ones by adapting advanced neural network architectures~\citep{ReconFormer,FPS,r8,aghabiglou2024mr,velayudham2025improving}. In other words, the main purpose of this methodology is to achieve MRI reconstruction purely in the image domain. Although they have succeeded in removing artifacts in the image domain, k-space recovery cannot be achieved, resulting in unsatisfactory reconstruction results.

Essentially, the aliasing artifacts arise because the accelerated undersampling destroys the k-space’s integrity. To this end, several works consider recovering k-space via an external network as shown in Figure.\ref{Fig.1} (a)~\citep{r35, r37,MD-Recon-Net,r38}. Despite some progress, they neglect the unique properties inherent to the k-space and directly adopt general CNNs which are not suitable for k-space recovery. CNNs have an inductive bias that their kernels are shared across spatial positions. However, in k-space, spatial positions represent frequency components of sine and cosine functions, and identical patterns at different positions can represent entirely different information\citep{r20,r25}. Previous works overlook this specific characteristic and adopt general image networks for k-space processing, resulting in inferior results. Hence, designing a customized network for undersampled k-space recovery is the key to reconstructing satisfactory MRI images.

Intuitively, a high-quality MRI image is generated from a continuous k-space where each coordinate has a specific signal. Therefore, the ideal reconstruction of k-space should be modeled as a function where arbitrary coordinates can be mapped into a k-value. Such a concept of continuous representation is consistent with implicit neural representation (INR), which emerged as a new paradigm for image super-resolution~\citep{r21,r22}. INR uses a neural network to model images as continuous functions of spatial coordinates instead of discrete pixel grids. Specifically, the neural network takes low-resolution coordinates as input and predicts corresponding pixel values, thereby constructing a latent space. After that, the model generates high-resolution images by querying dense coordinates from the latent space, preserving fine details and enabling arbitrary resolution output~\citep{r21,r23}. 

Inspired by this, we propose to boost the performance of MRI reconstruction by recovering complete k-space via INR. However, simply introducing INR to MRI reconstruction still faces several challenges. (1) INR originates from the processing of natural images, where rich structure information is provided for effective latent space modeling. Compared to it, the k-space spectrum exhibits a simple structure, making such continuous recovery challenging. Therefore, more information needed to be provided to guide k-space reconstruction. (2) The practice of accelerated scanning leads to excessively sparse sampled points, directly recovering a dense k-space from such sparsely sampled points inevitably introduces over-smoothing and distortion. A more gentle recovery paradigm should be designed to progressively reconstruct the dense k-space.

Based on these, we propose an Image-domain Guided K-space Recovery Network, named IGKR-Net. As shown in Figure.\ref{Fig.1} (b), we customize a k-space net from a new perspective of INR with image guidance for MRI reconstruction.

Specifically, (1) We propose an INR-based encoder-decoder structure, directly processing signals in the k-space domain. To detail, we design an encoder based on the standard transformer including multi-head self-attention (MSA) and feed-forward network (FFN), to encode the sampled k-values and their corresponding coordinates. Thus, a continuous latent space of the k-space spectrum is learned. Subsequently, we adopt standard transformer decoder layers consisting of multi-head cross-attention (MCA), MSA, and FFN, to dynamically query the unsampled k-values using their coordinates and thus reconstruct the k-space spectrum with high quality. (2) We design an image domain guidance module to mine image-domain information from the undersampled MRI images and guide the above k-space recovery. (3) We propose a tri-attention refinement module to further provide more details in the image domain. (4) To mitigate the issues of over-smoothing and distortion caused by directly recovering dense k-space from sparsely sampled points, we introduce a multi-stage training strategy to obtain finer reconstruction results step-by-step.

The main contributions of our work are as follows: 
\begin{itemize}
\item {An implicit neural representation based encoder-decoder structure is proposed to directly process signals in the k-space. Benefiting from this structure, our work can reconstruct the high-quality k-space.}

\item {An image domain guidance module is designed to mine information from undersampled MRI images. Due to the design of this, we provide rich information for the challenging k-space recovery.}

\item {A multi-stage training strategy is introduced to progressively reconstruct MRI images from low resolution to high resolution. Benefiting from this, we mitigate concerns of over-smoothing and distortion and obtain precise results.}

\item {Extensive experiments are conducted on both single-coil and multi-coil datasets, demonstrating the superiority of our method over previous approaches.}
\end{itemize} 

\section{Related Work}
\subsection{Deep Learning for Accelerated MRI}

Traditional MRI involves the acquisition of dense k-space, resulting in long scan time. In recent years, many deep neural networks have been explored for accelerated MRI. The pipeline for these methods usually consists of three steps, i.e., roughly recovering the complete k-space from undersampled data via zero-filling, generating low-quality MRI images via inverse fast Fourier transform (IFFT), and learning the mapping relationship between such low-quality MRI images and their corresponding high-quality ones by adapting advanced CNN architectures, including UNet~\citep{r28}, deep residual network~\citep{r29}, generative adversarial network~\citep{r30}, and deep cascaded network~\citep{r31}. 

Besides, benefiting from the capability of learning global information, transformers~\citep{Yang_2024_CVPR,yang2024tacklingambiguityperspectiveuncertainty,MoRe} recently achieved satisfactory performance for accelerated MRI. Feng et al.~\citep{r32,r33} pioneered the introduction of the transformer in the field of MRI reconstruction, integrating MRI reconstruction with super-resolution in a multitask fashion. Huang et al.~\citep{r34} proposed a parallel imaging coupled swin transformer-based model. Reconformer ~\citep{ReconFormer} introduced a locally pyramidal but globally columnar architecture to perceive multi-scale representation at any stage for MRI image reconstruction. To sum up, the data-driven nature of deep learning allows these methods to learn the mapping between low-quality input and artifact-free images. However, such networks operate only in the image domain, making it challenging to ensure the consistency of the k-space.

To this end, several attempts have been made to incorporate physics information in this line of work, including enforcing k-space consistency directly after image enhancement or adding k-space consistency as an additional cost function term during training. For example, Hyun et al.~\citep{r11} enhanced k-space consistency by directly replacing the generated k-values with the corresponding original ones. Yang et al.~\citep{r12} considered incorporating an additional k-space consistency term into the loss function. Though effective at reducing artifacts, they fundamentally only focus on image-domain restoration, with k-space information used for coarse correction or consistency supervision. The images relied upon by such methods are severely compromised, and this damage is irreversible~\citep{r26,r27}. Indeed, the continuity of the k-space is crucial for the recovery of high-quality MRI images. However, these approaches neglected this and resulted in inferior results.

\subsection{K-space Recovery}
Recently, some studies began to recover k-space. Taejoon Eo et al.~\citep{r35} proposed KIKINet, using a combination of 4 different CNNs to operate on k-space, image, k-space, and image sequentially. Osvald Nitski et al.~\citep{r36} proposed CDF-Net, using two UNet to operate on the image and k-space domain respectively, and combining them into an end-to-end framework. Wang et al.~\citep{10606502} introduced a compressed sensing equivariant imaging prior framework that combines a data preparation strategy for generalizable MRI reconstruction. Zhao et al.~\cite{r38} proposed SwinGAN, consisting of a k-space generator and an image-domain generator that utilizes swin transformer as the backbone. These approaches aimed to preserve and leverage the k-space information, achieving accurate results. However, they treated the k-space spectrums as natural images, directly employing general networks for feature extraction and spectrum reconstruction.

Besides, as for k-space based parallel reconstruction, GRAPPA~\citep{r56}, SPIRiT~\citep{r57} and AC-LORAKS~\citep{r58} used linear convolution to interpolate undersampled k-space. They assumed that the k-space data has a shifted autoregressive structure and learned the scan-specific autoregressive relationships to recover missing sample scans from the fully sampled auto-calibration (ACS) data. However, at high acceleration, they produce inherent noise amplification. Recently, the structure of the linear GRAPPA method has been transformed into a nonlinear deep learning method called RAKI~\citep{r59}. However, the availability of training ACS databases with different scans hinders its adaptability.

\begin{figure*}[!t]
\centerline{\includegraphics[width=1.0\linewidth]{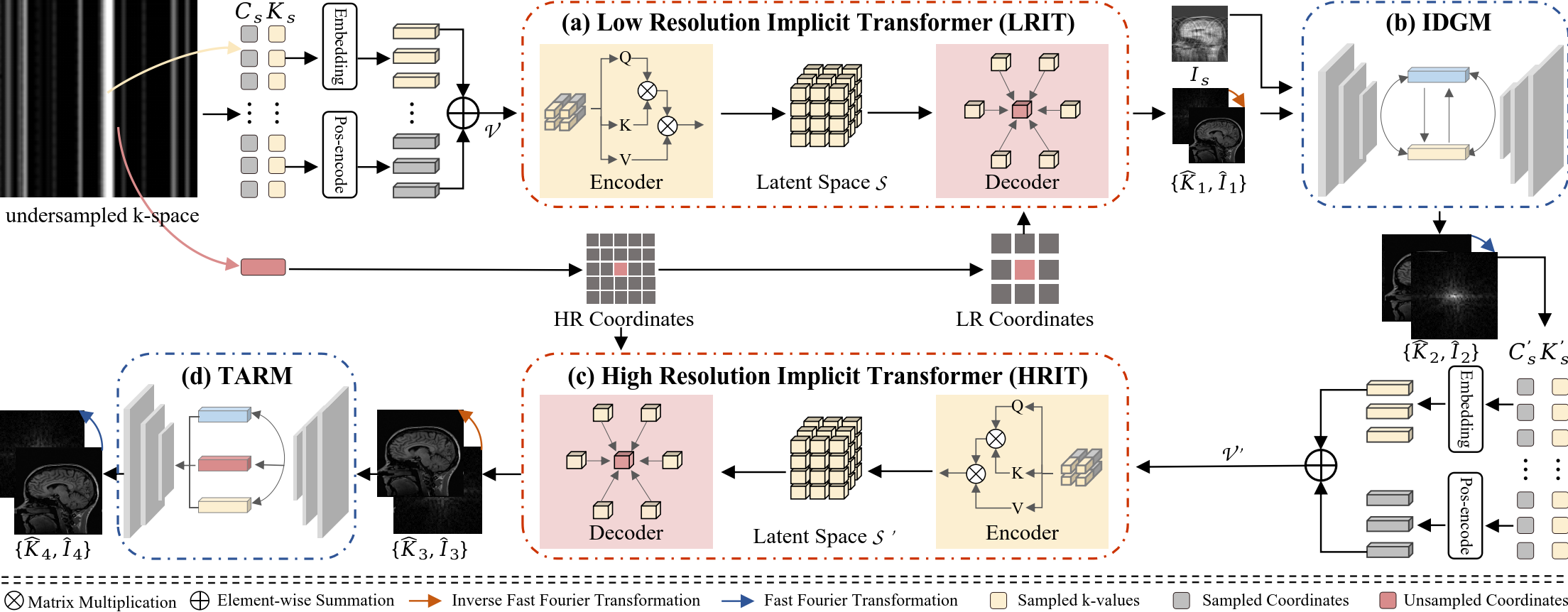}}
\caption{The architecture of the our IGKR-Net. Given the undersampled k-space, we first encode the coordinates $C_{s}$ and k-values $K_{s}$ of the sampled points, and the resulting features are added and then fed into LRIT. In LRIT, we use the LR coordinates to query and get the recovered results, i.e., $\widehat{{K}}_{1}$ and $\widehat{{I}}_{1}$. Next, we send $\widehat{{I}}_{1}$ into IDGM and get $\widehat{{K}}_{2}$ and $\widehat{{I}}_{2}$ with the guidance of the low-quality image $I_{s}$. Then, the k-values $K_{s}^{'}$ and coordinates $C_{s}^{'}$ of $\widehat{{K}}_{2}$ are encoded and fed into HRIT, where $\widehat{{K}}_{3}$ and $\widehat{{I}}_{3}$ are obtained by querying HR coordinates. Finally, we design TARM to refine $\widehat{{I}}_{3}$, yielding the final output $\widehat{{K}}_{4}$ and $\widehat{{I}}_{4}$.}
\vspace{-0.2 cm}
\label{Fig.2}
\end{figure*}
\begin{figure}[!t]
\centerline{\includegraphics[width=1.0\linewidth]{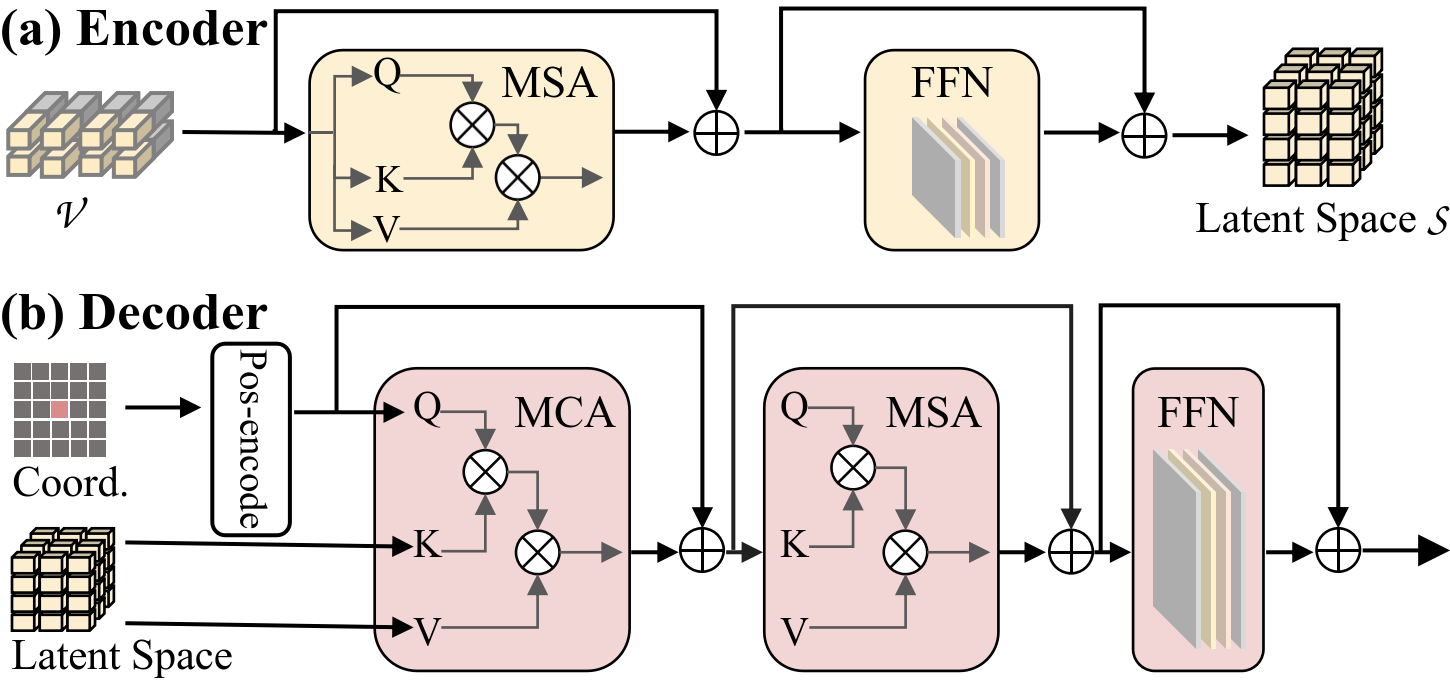}}
\caption{Our transformer based Encoder (a) and Decoder (b).}
\vspace{-0.4 cm}
\label{Fig.3}
\end{figure}
\subsection{Implicit Neural Representation}
Recently, implicit neural representation (INR) has been proposed as a method of continuous representation for various tasks~\citep{r39,r40,r41,r42,r43,r44}. INR employs neural networks, typically coordinate-based multilayer merceptrons (MLPs), to establish mapping relationships between coordinates and their corresponding signal values~\citep{10621671,10261266,LI2025126487}. 

Because of its continuous and precise representation, INR has been applied to various tasks of image enhancement~\citep{r21,r22}. For example, LIIF~\citep{r21} proposed a new framework for arbitrary-scale super-resolution using INR. Li et al.~\citep{r23} proposed a novel adaptive local image function (A-LIIF) to alleviate structural distortions and ringing artifacts around edges. Motivated by the feature interpolation in LIIF, Wu et al.~\citep{r51} designed an INR-based network for 3D MRI super-resolution at arbitrary scales. In summary, 2D INR models an image as a continuous function using latent space, enabling precise querying of signal intensity for any given coordinates. This concept provides a novel approach for achieving high-quality recovery in k-space.

\section{Methodology}
\subsection{Preliminary}
\subsubsection{MRI Reconstrucion}
Let $K$ represent the complex-valued, fully sampled k-space acquired from the MRI scanner, accelerated MRI usually employs undersampling to acquire a reduced set of k-space, i.e., $K_{s}$, while a substantial portion of the k-space, i.e., $K_{us}$, is unsampled. Here, we simulate this undersampling process by the element-wise multiplication ($\otimes$) of $K$ with a two-dimensional mask $M$ as follows:
\vspace{-0.1cm}
\begin{equation} 
K_{s}=M\otimes K,K_{us}=(1-M)\otimes K.
\label{eq1}
\end{equation}

Low-quality MRI image $I_{s}$ can be obtained, i.e., $I_{s} = IFFT(K_{s})$. To recover high-quality MRI image $\widehat{{I}}$, deep learning methods typically leverage extensive training data to establish a mapping relationship between $\widehat{{I}}$ and the sampled data $(I_{s}, K_{s})$, which can be formulated as follows:
\vspace{-0.1cm}
\begin{equation}
\widehat{{I}}=f_{\theta}(I_{s},K_{s}),
\label{eq2}
\end{equation}
where $\theta$ is the parameters set of the deep neural network. Different from these methods, our IGKR-Net focuses on recovering $K_{us}$, which is the fundamental cause of MRI image blurring and is crucial for high-quality MRI reconstruction.

\subsubsection{Implicit Neural Representation}
In contrast to the typical representation of images using discrete pixel arrays, implicit neural representation (INR) models the mapping relationship between image coordinates $C$ and their corresponding intensity values $i$ as a function, which can be formulated as follows:
\vspace{-0.1cm}
\begin{equation}
i = f_{\theta}(C,V_{c}),\
\label{eq3}
\end{equation}
where $V_{c}$ represents the feature vector specific to the pixel at spatial coordinates $C$. $V_{c}$ is obtained by establishing a latent space by training an encoder. Function $f_{\theta}$ is fitted using a neural network and is continuous to the spatial coordinate system. Therefore, once $f_{\theta}$ is well-fitted, intensity values can be queried at arbitrary coordinates, thereby achieving a continuous visual representation of the image.

\subsubsection{Continuous K-space Recovery Network via INR}
As described in equation \eqref{eq1}, it is common to undersample the k-space for accelerated MRI. The correct recovery of the unsampled k-values, i.e., $K_{us}$, is crucial for high-quality MRI reconstruction. Inspired by the continuous image modeling of INR, we customize a k-space recovery net. Specifically, $K_{us}$ can be queried via their coordinates $C_{us}$:
\vspace{-0.1cm}
\begin{equation}
K_{{us}}=f_{\theta}(C_{{us}},V_{C_{us}}),
\label{eq4}
\end{equation}
where the latent feature vectors $V_{C_{us}}$ are obtained by encoding the sampled k-values $K_{s}$ and their coordinates $C_{s}$ using a designed encoder $\Phi_{Enc}$. The function $f_{\theta}$ is fitted through training a decoder $\Phi_{Dec}$, which can be formulated as:
\vspace{-0.1cm}
\begin{equation}
K_{{us}}=\Phi_{Dec}(C_{{us}},\Phi_{Enc}(K_{s},C_{s})),
\label{eq5}
\end{equation}
where the encoder $\Phi_{Enc}$ and decoder $\Phi_{Dec}$ are implemented through the transformer attention mechanism.

\subsection{Network Architecture}
As shown in Figure.\ref{Fig.2}, the overall architecture of our IGKR-Net can be divided into two parts, i.e., the primary part directly processing in k-space domain (denoted as \textcolor{red}{red dash lines}), and the auxiliary part operating on image domain (denoted as \textcolor{blue}{blue dash lines}). The k-space part comprises LRIT and HRIT, to reconstruct the k-space from low resolution (LR) to high resolution (HR) progressively, while the image part comprises IDGM and TRAM for image guidance and final refinement, respectively.

Specifically, given undersampled k-space $K \in  R^{2\times H \times W}$ as input, we obtain the set of k-values $K_{s}$ and the corresponding coordinates $C_{s}$, from the sampled points. Then, $K_{s}$ and $C_{s}$ are encoded separately, and the resulting features are summed to form a vector sequence $\mathcal{V}$, which is used as the input of LRIT. LRIT first uses a transformer-based encoder to learn a latent space $\mathcal{S}$, and then recover a complete while LR k-space $\widehat{{K}}_{1} \in  R^{2\times \frac{H}{2} \times \frac{W}{2}}$ by querying pre-designed LR coordinates through the decoder. Subsequently, we get $\widehat{{I}}_{1}$ through IFFT and send it into IDGM to deeply mine information from the corrupted image $I_{s}$ and guide the k-space recovery. We fuse $I_{s}$ with $\widehat{{I}}_{1}$, obtaining $\widehat{{I}}_{2}\in  R^{2\times H \times W}$ and $\widehat{{K}}_{2}\in  R^{2\times H \times W}$. Next, HRIT uses the recovered dense $\widehat{{K}}_{2}$ to further establish a continuous latent space $\mathcal{S}^{'}$, recovering complete and HR $\widehat{K}_{3}$ and its corresponding $\widehat{{I}}_{3}$ using the HR coordinates. Finally, we design TARM to refine $\widehat{{I}}_{3}$ in the image domain, resulting in the final output, i.e., $\widehat{{I}}_{4}$ and $\widehat{{K}}_{4}$. 

\begin{figure}[!t]
\centerline{\includegraphics[width=1.0\linewidth]{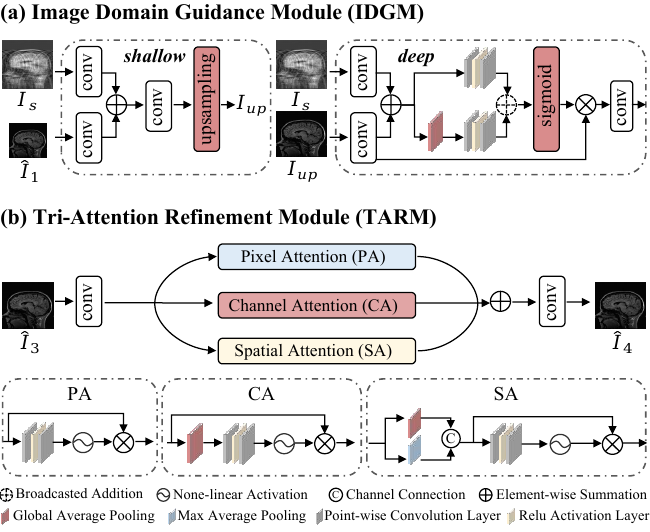}}
\caption{(a) The proposed IDGM consists of two stages, i.e., shallow fusion stage and deep fusion stage. (b) The proposed TARM consists of three branches, i.e., PA, CA, and SA.}
\vspace{-0.5 cm}
\label{Fig.4}
\end{figure}
Besides, to mitigate issues of over-smoothing and distortion and obtain satisfactory results progressively, a multi-stage training strategy is proposed. Below we first present the design of LRIT, IDGM, HRIT, and TARM, respectively, followed by a description of the training strategy.

\subsubsection{Low Resolution Implicit Transformer (LRIT)}
As depicted in Figure.\ref{Fig.2} (a), LRIT is an encoder-decoder architecture. It aims to recover LR k-space $\widehat{{K}}_{1}$ and its corresponding $\widehat{{I}}_{1}$ by querying the LR coordinates.

Specifically, given a set of sampled k-values $K_{s}$ and its coordinates $C_{s}$, a tokenization procedure is firstly applied to convert them into a vector sequence $\mathcal{V}$ as follows:
\vspace{-0.1cm}
\begin{equation}
\mathcal{V} = EM(K_{s})+PE(C_{s}),
\label{eq6}
\end{equation}
where $EM$ is the embedding operation via multilayer perceptrons (MLPs) layer, and $PE$ refers to positional encodings with sine and cosine functions. Then we send $\mathcal{V}$ into the encoder of LRIT, which consists of multi-head self-attention ($MSA$) and feed-forward network ($FFN$) as shown in Figure.\ref{Fig.3} (a). There by, we can obtain the latent space $\mathcal{S}$ as:
\vspace{-0.1cm}
\begin{equation}
\mathcal{S} = FFN(MSA(\mathcal{V})).
\label{eq7}
\end{equation}

Subsequently, we design a decoder that employs standard transformer decoder layers consisting of multi-head cross attention (MCA), MSA, and FFN as shown in Figure.\ref{Fig.3} (b), to recover complete while low-resolution k-space, i.e., $\widehat{{K}}_{1}$. In detail, for MCA, we employ encoded LR coordinates ${C}_{lr} \in R^{\frac{H}{2} \times \frac{W}{2}}$ to serve as the query ($Q$) and apply two different linear projects to latent space $\mathcal{S}$ to serve as the key ($K$) and value ($V$). After that, we apply MSA to the weighted feature again and decode it via FFN. Thus, we obtain a low-resolution while complete k-space $\widehat{{K}}_{1}\in R^{2\times \frac{H}{2} \times \frac{W}{2} }$, as well as its corresponding MRI image $\widehat{{I}}_{1}\in R^{1\times \frac{H}{2} \times \frac{W}{2} }$: 
\vspace{-0.1cm}
\begin{equation}
\widehat{{K}}_{1}=FFN(MSA(MCA(Q,K,V)),
\label{eq8}
\end{equation}
\vspace{-0.5cm}
\begin{equation}
\widehat{{I}}_{1}={IFFT}(\widehat{{K}}_{1}).
\label{eq9}
\end{equation}
\vspace{-0.8cm}
\begin{algorithm}[!t]
\caption{Training procedure}
\LinesNumbered
{$stage\_1$: Train LRIT}\;
\For{ each $epoch\in \left [ E_{0},E_{1}\right]$ }
{
    Get $({{K_{s}}},{{C_{s}}},{C_{lr}},{K_{lr}},{I_{lr}})$;\\
    Train LRIT by minimizing \\
            $L_{stage\_1} = L_{1} $\
}
{$stage\_2$: Joint optimization of LRIT and DIFM}\;
\For{ each $epoch\in \left [ E_{1},E_{2}\right]$ }
{
    Get $({{K_{s}}},{{C_{s}}},{C_{lr}},{K_{lr}},{I_{lr}},{I_{s}},{K},{I})$;\\
    Train LRIT and DIFM by minimizing \\
            $L_{stage\_2} = L_{1} + L_{2} $\
}
{$stage\_3$: Joint optimization of LRIT, DIFM, and HRIT}\;
\For{ each $epoch\in \left [ E_{2},E_{3}\right]$ }
{
    Get $({{K_{s}}},{{C_{s}}},{C_{lr}},{K_{lr}},{I_{lr}},{I_{s}},{K},{I},{C_{us}})$;\\
    Train LRIT, DIFM, and HRIT by minimizing \\
            $L_{stage\_3} = L_{1} + L_{2} + L_{3} $\
}
{$stage\_4$: Joint optimization of LRIT, DIFM, HRIT, and TARM}\;
\For{ each $epoch\in \left [ E_{3},E_{4}\right]$ }
{
    Get $({{K_{s}}},{{C_{s}}},{C_{lr}},{K_{lr}},{I_{lr}},{I_{s}},{K},{I},{C_{us}})$;\\
    Train LRIT, DIFM, HRIT, TARM by minimizing \\
            $L_{stage\_4} = L_{1} + L_{2} + L_{3} + L_{4}$\
}
\label{A1}
\end{algorithm}


\subsubsection{Image Domain Guidance Module (IDGM)}
As shown in Figure.\ref{Fig.2} (b), our IDGM fuses the low-quality image $I_{s}$ and the output of LRIT, i.e., $\widehat{{I}}_{1}$, to fully utilize the image information and guide the k-space recovery.

Figure.\ref{Fig.4} (a) illustrates the detailed structure of the IDGM, which can be divided into two stages. Specifically, during the shallow stage, we firstly perform two convolutions with different strides to extract shallow features from $I_{s}$ and $\widehat{{I}}_{1}$ and unify them to the same size, i.e., ${\widehat {I}_{1}} \in  R^{1\times \frac{H}{2} \times \frac{W}{2} }\to  R^{h\times \frac{H}{2} \times \frac{W}{2}}$, ${I_{s}} \in  R^{1\times H \times W }\to  R^{h\times \frac{H}{2} \times \frac{W}{2}}$, where h denotes the channel number. Then, the features are fused via element-wise summation and restored to ${{I}}_{up} \in  R^{1\times H \times W }$ via convolution $conv(\cdot)$ and upsampling $up(\cdot)$ as follows:
\vspace{-0.1cm}
\begin{equation}
I_{up} = up(conv(conv({\widehat {{I}}_{1}})+conv(I_{s} ))).
\label{eq10}
\end{equation}

During the deep fusion stage, we first employ two convolution heads to extract image-specific shallow features from $I_{s}$ and ${{I}}_{up}$, respectively. After element-wise summation, the fused feature is sent into two attention branches, generating local channel attention $A_{l}$ without any pooling and global channel attention $A_{g}$ by using global average pooling, respectively. Given $A_{l}$ and $A_{g}$, the final attention weights $A$ can be obtained by the broadcasted addition ($\textbf{+}$), i.e., $A= sigmoid(A_{l} \textbf{+} A_{g})$. Finally, we obtain $\widehat{{I}}_{2}$, as well as its corresponding k-space $\widehat{{K}}_{2}$ as follows:
\vspace{-0.1cm}
\begin{equation}
\widehat {{I}}_{2} = conv(A\otimes conv({I}_{up})),
\label{eq11}
\end{equation}
\vspace{-0.5cm}
\begin{equation}
\widehat{{K}}_{2}={FFT}(\widehat{{I}}_{2}).
\label{eq12}
\end{equation}

\subsubsection{High Resolution Implicit Transformer (HRIT)}
As shown in Figure.\ref{Fig.2} (c), to further reconstruct the dense HR k-space, we apply the implicit transformer again. Different from LRIT, here we use the recovered dense $\widehat{{K}}_{2}$ to generate $K_{s}'$ and $C_{s}'$, forming vector sequence $\mathcal{V'}$:
\vspace{-0.1cm}
\begin{equation}
\mathcal{V'} = EM(K_{s}')+PE(C_{s}').
\label{eq13}
\end{equation}
Then we get latent space $\mathcal{S'}$ via the Encoder in Figure.\ref{Fig.4} (a):
\vspace{-0.1cm}
\begin{equation}
\mathcal{S'} = FFN(MSA(\mathcal{V'})).
\label{eq14}
\end{equation}
Finally, we encoded HR coordinates $C_{hr}\in  R^{ H \times W }$ as query ($Q'$) and projecting $\mathcal{S'}$ as key ($K'$) and value ($V'$). By using the Decoder shown in Figure.\ref{Fig.4} (b), we can obtain the high quality k-space $\widehat{{K}}_{3}$ and its corresponding  MRI image $\widehat{{I}}_{3}$:
\vspace{-0.1cm}
\begin{equation}
\widehat{{K}}_{3}=FFN(MSA(MCA(Q',K',V')),
\label{eq15}
\end{equation}
\vspace{-0.5cm}
\begin{equation}
\widehat{{I}}_{3}={IFFT}(\widehat{{K}}_{3}).
\label{eq16}
\end{equation}
\vspace{-0.8cm}
\renewcommand{\dblfloatpagefraction}{.9}
\begin{figure*}[!t]
\centerline{\includegraphics[width=1.0\linewidth]{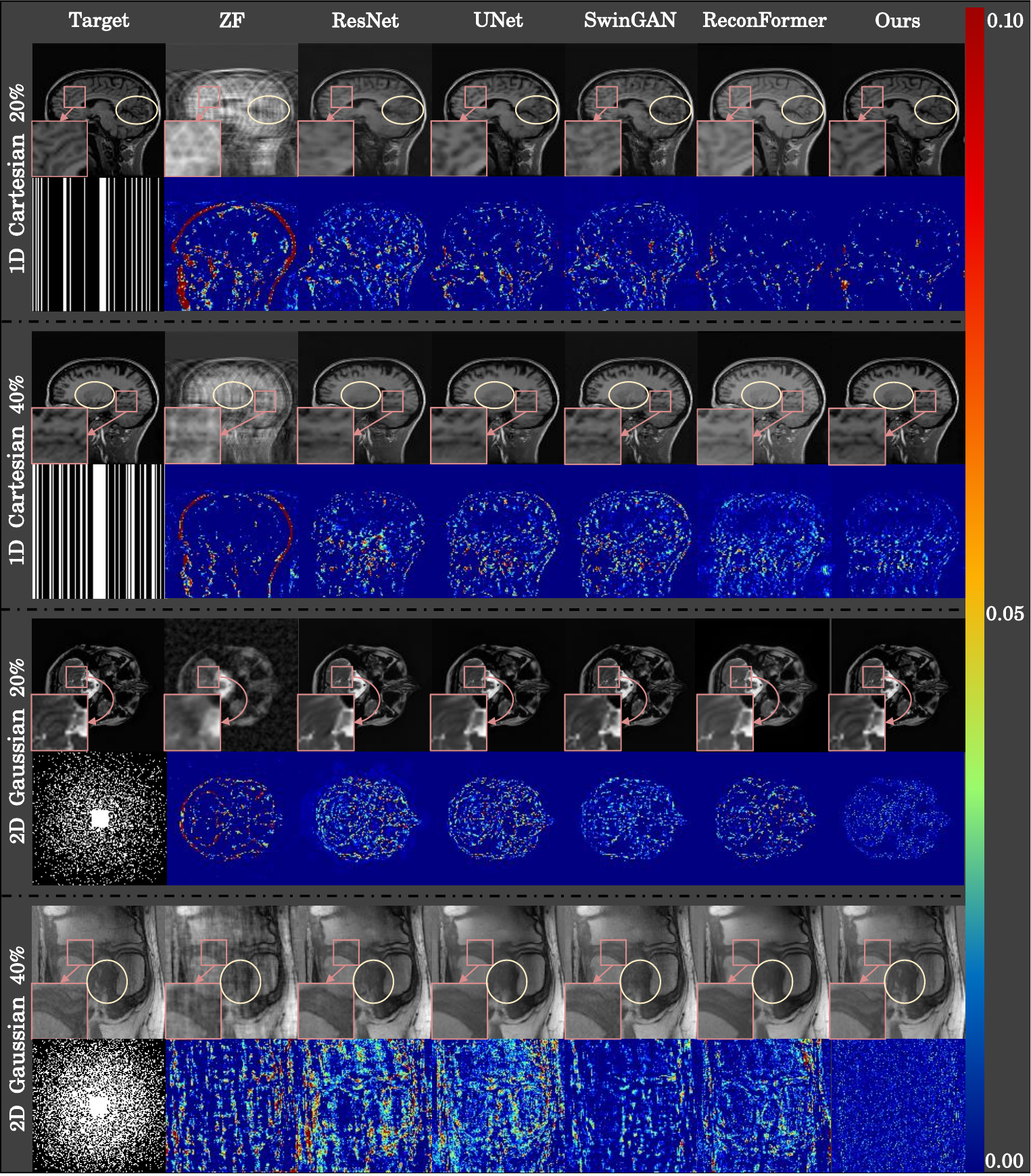}}
\caption{Visual comparison of different methods under various undersampling masks. Red boxes illustrate the enlarged views in detail. Yellow ellipses highlight the performance differences in the results of anatomical structures among various methods.}
\label{Fig.5}
\vspace{-0.4cm}
\end{figure*}
\subsubsection{Tri-Attention Refinement Module (TARM)}
As shown in Figure.\ref{Fig.2} (d), our TARM refines the reconstructed $\widehat{{I}}_{3}$ in image domain, providing it with richer details.

Figure.\ref{Fig.4} (b) illustrates the structure of the TARM, which comprises a convolution head and three attention branches, i.e., spatial attention (SA), channel attention (CA), and pixel attention (PA). Specifically, for an image that needs to be refined, i.e., $\widehat{{I}}_{3}$, the TARM first extracts shallow feature $f$ by a convolution. Then, $f$ is sent into three attention branches. To detail, PA generates a 3D attention mask $M_{1}$ without any pooling or sampling, which means the output features have local information. CA uses global average pooling (GAP) to generate the 1D attention mask, i.e., $M_{2}$, to obtain features that have global information. SA generates a 2D attention mask, i.e., $M_{3}$, to get features with global information. Finally, the weighted features from three attention branches are fused via element-wise summation, to obtain the final results $\widehat{{I}}_{4}$ and its corresponding k-space $\widehat{{K}}_{4}$ as follows:
\vspace{-0.1cm}
\begin{equation}
\widehat{{I}}_{4}= conv(M_{1}\otimes f +M_{2}\otimes f+M_{3}\otimes f),
\label{eq17}
\end{equation}
\vspace{-0.5cm}
\begin{equation}
\widehat{{K}}_4={FFT}(\widehat{{I}}_4).
\label{eq18}
\end{equation}
\begin{table*}[!t]
\renewcommand{\arraystretch}{1.1}
    \centering
    \small
    \setlength{\tabcolsep}{0.9mm}
    \caption{Quantitative results of the comparison experiments on the CC359 validation dataset using different masks. A higher value of metric↑ indicates better performance, whereas a lower value of metric↓ corresponds to improved performance.}
\begin{tabular}{l|cccc|cccc}
\toprule
\multicolumn{1}{c|}{}                         & \multicolumn{4}{c|}{1D Cartesian mask, sampling ratio = 20\%}                               & \multicolumn{4}{c}{1D Cartesian mask, sampling ratio = 40\%}                                \\ \cmidrule{2-9} 
\multicolumn{1}{c|}{\multirow{-2}{*}{Method}} & PSNR↑                & SSIM↑                 & NMSE↓                 & LPIPS↓                & PSNR↑                & SSIM↑                 & NMSE↓                 & LPIPS↓                \\ \midrule
ZF                                            & 21.62±2.80          & 0.6254±0.11          & 0.0936±0.03          & 0.3596±0.04          & 22.74±2.83          & 0.6784±0.08          & 0.0723±0.02          & 0.2969±0.04          \\
ResNet                                        & 28.13±3.84          & 0.8559±0.14          & 0.0200±0.01          & 0.0858±0.07          & 30.05±3.58          & 0.9052±0.07          & 0.0131±0.01          & 0.0569±0.04          \\
UNet                                         & 29.34±4.59          & 0.8756±0.11          & 0.0156±0.01          & 0.0618±0.05          & 31.73±4.70          & 0.9189±0.07          & 0.0090±0.00          & 0.0408±0.04          \\
SwinMR                                       & 31.11±3.87          & 0.9006±0.09          & 0.0124±0.00          & 0.0413±0.03          & 35.67±4.13          & 0.9543±0.04          & 0.0045±0.00          & 0.0184±0.02          \\
RefineGAN                                     & 29.96±4.59          & 0.8888±0.07          & 0.0303±0.02          & 0.0395±0.04          & 38.57±5.16          & 0.9656±0.02          & 0.0058±0.01          & 0.0077±0.01          \\
SwinGAN                                       & 31.32±3.20          & 0.9017±0.06          & 0.0087±0.01          & 0.0390±0.02          & 38.67±4.56          & 0.9688±0.02          & 0.0043±0.01          & 0.0156±0.02          \\
ReconFormer                                   & 32.79±2.38          & 0.9189±0.05          & 0.0072±0.02          & 0.0373±0.03          & 38.75±4.77          & 0.9696±0.02          & 0.0032±0.03          & 0.0149±0.02          \\
\rowcolor[HTML]{EFEFEF} 
\textbf{Ours}                                 & \textbf{33.06±5.27} & \textbf{0.9234±0.10} & \textbf{0.0069±0.00} & \textbf{0.0362±0.04} & \textbf{38.78±5.67} & \textbf{0.9703±0.05} & \textbf{0.0020±0.00} & \textbf{0.0143±0.03} \\ \midrule
                                              & \multicolumn{4}{c|}{2D Gaussian mask, sampling ratio = 20\%}                               & \multicolumn{4}{c}{2D Gaussian mask, sampling ratio = 40\%}                                \\ \midrule
ZF                                            & 24.23±3.18          & 0.5942±0.10          & 0.0497±0.02          & 0.3796±0.06          & 27.22±3.39          & 0.6901±0.07          & 0.0246±0.01          & 0.2543±0.08          \\
ResNet                                       & 28.83±3.47          & 0.8576±0.12          & 0.0185±0.02          & 0.0841±0.08          & 31.00±3.66          & 0.9102±0.09          & 0.0181±0.05          & 0.0592±0.10          \\
UNet                                          & 30.39±4.78          & 0.8790±0.12          & 0.0124±0.00          & 0.0578±0.05          & 33.39±4.98          & 0.9280±0.08          & 0.0063±0.00          & 0.0362±0.03          \\
SwinMR                                        & 31.44±4.01          & 0.8943±0.09          & 0.0116±0.04          & 0.0441±0.03          & 34.36±3.85          & 0.9378±0.05          & 0.0059±0.00          & 0.0220±0.01          \\
RefineGAN                                     & 30.08±4.61          & 0.8770±0.07          & 0.0303±0.02          & 0.0395±0.03          & 35.73±5.05          & 0.9486±0.03          & 0.0098±0.01          & 0.0140±0.01          \\
SwinGAN                                       & 34.56±3.56          & 0.9233±0.04          & 0.0078±0.01          & 0.0221±0.02          & 38.89±4.32          & 0.9676±0.04          & 0.0046±0.02          & 0.0089±0.01          \\
ReconFormer                                    & 36.01±3.35          & 0.9400±0.04          & 0.0045±0.02          & 0.0192±0.01          & 41.35±2.78          & 0.9784±0.04          & 0.0022±0.01          & 0.0055±0.01          \\
\rowcolor[HTML]{EFEFEF} 
\textbf{Ours}                                 & \textbf{37.09±5.54} & \textbf{0.9512±0.09} & \textbf{0.0029±0.00} & \textbf{0.0184±0.03} & \textbf{43.04±5.72} & \textbf{0.9807±0.05} & \textbf{0.0008±0.00} & \textbf{0.0048±0.01} \\ \bottomrule
\end{tabular}
    \label{table.1}
\end{table*}

\begin{table*}[!h]
\renewcommand{\arraystretch}{1.1}
    \centering
    \small
    \setlength{\tabcolsep}{0.9mm}
    \caption{Quantitative results of the comparison experiments on the fastMRI validation dataset using different masks. A higher value of metric↑ indicates better performance, whereas a lower value of metric↓ corresponds to improved performance.}
\begin{tabular}{l|cccc|cccc}
\toprule
\multicolumn{1}{c|}{}                         & \multicolumn{4}{c|}{1D Cartesian mask, sampling ratio = 20\%}                               & \multicolumn{4}{c}{1D Cartesian mask, sampling ratio = 40\%}                                \\ \cmidrule{2-9} 
\multicolumn{1}{l|}{\multirow{-2}{*}{Method}} & PSNR↑                & SSIM↑                 & NMSE↓                 & LPIPS↓                & PSNR↑                & SSIM↑                 & NMSE↓                 & LPIPS↓                \\ \midrule
ZF                                            & 21.96±2.19          & 0.5022±0.07          & 0.0818±0.03          & 0.4237±0.06          & 23.04±2.28          & 0.6011±0.06          & 0.0638±0.02          & 0.3444±0.06          \\
ResNet                                        & 24.54±2.02          & 0.6062±0.09          & 0.0461±0.02          & 0.2469±0.08          & 25.75±2.04          & 0.7059±0.07          & 0.0351±0.02          & 0.1791±0.06          \\
UNet                                          & 26.53±2.22          & 0.6599±0.10          & 0.0316±0.02          & 0.2242±0.08          & 28.21±2.33          & 0.7559±0.07          & 0.0220±0.02          & 0.1541±0.06          \\
SwinMR                                        & 24.67±2.21          & 0.6214±0.10          & 0.0480±0.03          & 0.2166±0.06          & 27.33±2.46          & 0.7530±0.08          & 0.0271±0.02          & 0.1354±0.05          \\
RefineGAN                                     & 25.71±2.56          & 0.6961±0.06          & 0.0335±0.02          & 0.1922±0.04          & 29.38±2.37          & 0.8054±0.02          & 0.0174±0.08          & 0.1376±0.01          \\
SwinGAN                                       & 26.60±2.09          & 0.7001±0.08          & 0.0330±0.01          & 0.1996±0.03          & 29.89±2.33          & 0.8069±0.06          & 0.0169±0.07          & 0.1225±0.04          \\
ReconFormer                                    & 27.65±2.34          & 0.6999±0.04          & 0.0269±0.01          & 0.1887±0.04          & 30.03±2.54          & 0.8072±0.03          & 0.0154±0.03          & 0.1170±0.03          \\
\rowcolor[HTML]{EFEFEF} 
\textbf{Ours}                                 & \textbf{27.67±2.52} & \textbf{0.7011±0.11} & \textbf{0.0266±0.02} & \textbf{0.1843±0.07} & \textbf{30.07±2.87} & \textbf{0.8073±0.08} & \textbf{0.0152±0.07} & \textbf{0.1128±0.05} \\ \midrule
                                              & \multicolumn{4}{c|}{2D Gaussian mask, sampling ratio = 20\%}                               & \multicolumn{4}{c}{2D Gaussian mask, sampling ratio = 40\%}                                \\ \midrule
ZF                                            & 24.84±1.99          & 0.5840±0.07          & 0.0438±0.02          & 0.3364±0.06          & 27.57±2.03          & 0.7232±0.05          & 0.0242±0.02          & 0.1985±0.05          \\
ResNet                                        & 26.60±2.09          & 0.6611±0.10          & 0.0305±0.02          & 0.2097±0.06          & 27.91±2.04          & 0.7600±0.07          & 0.0227±0.01          & 0.1176±0.04          \\
UNet                                          & 28.10±2.47          & 0.6875±0.11          & 0.0236±0.02          & 0.2058±0.06          & 30.20±2.71          & 0.7869±0.07          & 0.0155±0.01          & 0.1094±0.04          \\
SwinMR                                        & 27.42±2.55          & 0.6813±0.11          & 0.0278±0.02          & 0.1994±0.05          & 29.54±2.68          & 0.7802±0.07          & 0.0179±0.01          & 0.0988±0.03          \\
RefineGAN                                    & 28.29±2.53          & 0.7766±0.06          & 0.0211±0.01          & 0.1533±0.03          & 31.52±2.41          & 0.8253±0.03          & 0.0150±0.01          & 0.0681±0.02          \\
SwinGAN                                       & 29.22±2.34          & 0.7455±0.10          & 0.0200±0.01          & 0.1502±0.04          & 31.89±2.44          & 0.8277±0.04          & 0.0144±0.02          & 0.0677±0.02          \\
ReconFormer                                    & 29.73±2.62          & 0.7485±0.04          & 0.0199±0.02          & 0.1514±0.02          & 32.02±1.89          & 0.8292±0.04          & 0.0156±0.02          & 0.0681±0.02          \\
\rowcolor[HTML]{EFEFEF} 
\textbf{Ours}                                 & \textbf{29.84±3.18} & \textbf{0.7500±0.11} & \textbf{0.0188±0.02} & \textbf{0.1497±0.06} & \textbf{32.07±3.65} & \textbf{0.8394±0.07} & \textbf{0.0124±0.01} & \textbf{0.0670±0.03} \\ \bottomrule
\end{tabular}
    \label{table.2}
\vspace{-1.0 em}
\end{table*}
\begin{table*}[]
\renewcommand{\arraystretch}{1.1}
    \centering
    \small
    \setlength{\tabcolsep}{0.9mm}
    \caption{Quantitative results of the comparison experiments on the IXI test dataset using different masks. A higher value of metric↑ indicates better performance, whereas a lower value of metric↓ corresponds to improved performance.}
\begin{tabular}{l|cccc|cccc}
\toprule
\multicolumn{1}{c|}{}                         & \multicolumn{4}{c|}{1D Cartesian mask, sampling ratio = 20\%}                               & \multicolumn{4}{c}{1D Cartesian mask, sampling ratio = 40\%}                                \\ \cmidrule{2-9} 
\multicolumn{1}{l|}{\multirow{-2}{*}{Method}} & PSNR↑                & SSIM↑                 & NMSE↓                 & LPIPS↓                & PSNR↑                & SSIM↑                 & NMSE↓                 & LPIPS↓                \\ \midrule
ZF                                            & 22.57±2.42          & 0.6994±0.07          & 0.1406±0.03          & 0.3158±0.04          & 23.81±2.46          & 0.7443±0.06          & 0.1058±0.03          & 0.2509±0.04          \\
ResNet                                        & 29.23±3.18          & 0.9249±0.04          & 0.0323±0.01          & 0.0416±0.02          & 32.69±3.27          & 0.9609±0.02          & 0.0148±0.01          & 0.0229±0.01          \\
UNet                                          & 30.54±3.43          & 0.9434±0.03          & 0.0244±0.01          & 0.0291±0.01          & 33.72±3.53          & 0.9685±0.02          & 0.0119±0.01          & 0.0159±0.01          \\
SwinMR                                        & 34.15±3.51          & 0.9652±0.02          & 0.0123±0.01          & 0.0133±0.01          & 37.70±3.57          & 0.9805±0.01          & 0.0055±0.00          & 0.0055±0.01          \\
RefineGAN                                     & 32.96±2.58          & 0.9500±0.01          & 0.0126±0.01          & 0.0142±0.01          & 40.33±3.22          & 0.9788±0.02          & 0.0039±0.01          & 0.0028±0.01          \\
SwinGAN                                       & 33.53±3.32          & 0.9633±0.02          & 0.0120±0.01          & 0.0145±0.01          & 41.42±3.43          & 0.9891±0.01          & 0.0031±0.00          & 0.0022±0.00          \\
ReconFormer                                    & 33.89±2.87          & 0.9658±0.02          & 0.0120±0.02          & 0.0139±0.01          & 41.19±2.76         & 0.9876±0.02          & 0.0027±0.01          & \textbf{0.0022±0.01} \\
\rowcolor[HTML]{EFEFEF} 
\textbf{Ours}                                 & \textbf{34.40±3.98} & \textbf{0.9710±0.02} & \textbf{0.0113±0.01} & \textbf{0.0132±0.01} & \textbf{41.48±4.09} & \textbf{0.9924±0.01} & \textbf{0.0023±0.00} & 0.0033±0.00          \\ \midrule
                                              & \multicolumn{4}{c|}{2D Gaussian mask, sampling ratio = 20\%}                               & \multicolumn{4}{c}{2D Gaussian mask, sampling ratio = 40\%}                                \\ \midrule
ZF                                            & 24.49±3.02          & 0.6020±0.09          & 0.0924±0.02          & 0.3774±0.05          & 27.76±3.01          & 0.6968±0.07          & 0.0436±0.01          & 0.2617±0.05          \\
ResNet                                        & 31.47±3.65          & 0.9446±0.03          & 0.0199±0.01          & 0.0326±0.02          & 34.98±3.58          & 0.9715±0.02          & 0.0088±0.00          & 0.0183±0.02          \\
UNet                                          & 33.19±4.07          & 0.9608±0.03          & 0.0141±0.01          & 0.0234±0.02          & 37.12±4.01          & 0.9809±0.01          & 0.0057±0.00          & 0.0135±0.01          \\
SwinMR                                        & 32.89±3.42          & 0.9420±0.02          & 0.0155±0.01          & 0.0172±0.01          & 35.73±3.28          & 0.9590±0.02          & 0.0080±0.00          & 0.0085±0.01          \\
RefineGAN                                     & 36.36±4.01          & 0.9703±0.02          & 0.0109±0.01          & 0.0073±0.01          & 44.80±3.91          & 0.9910±0.01          & 0.0017±0.00          & 0.0012±0.00          \\
SwinGAN                                       & 37.44±3.34          & 0.9800±0.02          & 0.0056±0.01          & 0.0067±0.01          & 45.67±4.43          & 0.9954±0.00          & 0.0009±0.00          & 0.0010±0.00          \\
ReconFormer                                    & 39.52±2.76          & 0.9857±0.03          & 0.0033±0.02          & 0.0064±0.02          & 46.65±3.23          & 0.9868±0.01          & 0.0010±0.02          & 0.0009±0.02          \\
\rowcolor[HTML]{EFEFEF} 
\textbf{Ours}                                 & \textbf{40.61±4.12} & \textbf{0.9888±0.01} & \textbf{0.0029±0.00} & \textbf{0.0056±0.00} & \textbf{47.83±4.18} & \textbf{0.9973±0.00} & \textbf{0.0006±0.00} & \textbf{0.0008±0.00} \\ \bottomrule
\end{tabular}
    \label{table.3}
\vspace{-1.0 em}
\end{table*}
\begin{table}[]
\renewcommand{\arraystretch}{1.1}
    \centering
    \small
    \setlength{\tabcolsep}{0.4mm}
    \caption{Performance comparison of MRI reconstruction under $25\%$ and $12.5\%$ 1D Cartesian mask on the multi-coil SKM-TEA.}
\begin{tabular}{l|ccc|ccc}
\toprule
\multicolumn{1}{c|}{}                          & \multicolumn{3}{c|}{sampling ratio = 25\%}         & \multicolumn{3}{c}{sampling ratio = 12.5\%}        \\ \cmidrule{2-7} 
\multicolumn{1}{l|}{\multirow{-2}{*}{Methods}} & PSNR↑           & SSIM↑            & NMSE↓            & PSNR↑           & SSIM↑            & NMSE↓            \\ \midrule
UNet                                        & 33.91          & 0.8469          & 0.0204          & 31.44          & 0.7904          & 0.0204          \\
KIKI-Net                                       & 34.26          & 0.8577          & 0.0196          & 31.42          & 0.7941          & 0.0196          \\
SwinMR                                         & 34.45          & 0.8597          & 0.0192          & 31.94          & 0.8022          & 0.0192          \\
D5C5                                           & 34.63          & 0.8648          & 0.0188          & 31.89          & 0.8030          & 0.0188          \\
ReconFormer                                    & 35.06          & 0.8730          & 0.0179          & 32.51          & 0.8158          & 0.0179          \\
\rowcolor[HTML]{EFEFEF} 
\textbf{Ours}                                  & \textbf{35.14} & \textbf{0.8832} & \textbf{0.0167} & \textbf{32.66} & \textbf{0.8198} & \textbf{0.0173} \\ \bottomrule
\end{tabular}
    \label{table4}
\vspace{-1.5 em}
\end{table}
\subsubsection{Loss Functions and Training Strategy}
Following \cite{r37,r38}, we evaluate the intermediate outputs of various modules by using the $L_{2}$ loss, which consists of two terms. To detail, the first term is used to constrain the consistency between the reconstructed k-space $\widehat{{K}}_{i\in \left \{ 1,2,3,4 \right \}}$ and corresponding fully sampled k-space, i.e., $K_{lr}$ for $\widehat{{K}}_{1}$ and $K$ for $\widehat{{K}}_{i\in \left \{ 2,3,4 \right \}}$. The second term is used to constrain the consistency between the reconstructed image $\widehat{{I}}_{i\in \left \{ 1,2,3,4 \right \}}$ and the corresponding fully sampled image, i.e., $I_{lr}$ for $\widehat{{I}}_{1}$ and $I$ for $\widehat{{I}}_{i\in \left \{ 2,3,4 \right \}}$:
\vspace{-0.1cm}
\begin{equation}
L_{i}=\begin{cases}
 \left \| K_{lr}-\widehat{K}_{i}\right \| _{2} + \left \| I_{lr}-\widehat{I}_{i}\right \| _{2}  & \text{ if }i=1 \\[11pt]
 \left \| K_{\textcolor{white}{lr}} -\widehat{K}_{i}\right \| _{2} + \left \| I_{\textcolor{white}{lr}}-\widehat{I}_{i}\right \| _{2} & \text{ if } i=2,3,4
\end{cases} 
\label{eq19}
\end{equation}

where $I$, $I_{lr}$, and $K_{lr}$ is obtained via $I=IFFT(K)$, $I_{lr}=downsample(I)$, and $k_{lr}=FFT(I_{lr})$, respsectively. Furthermore, to mitigate the issues of over-smoothing and distortion and obtain finer reconstruction results step-by-step, we design a multi-stage training strategy, as illustrated by Algorithm \ref{A1}. Specifically, we divide the training process into four stages, i.e., $stage=stage_{i}, when {\textcolor{white}{1}} epoch\in \left [ E_{i-1},E_{i}\right] |_{i=1}^{i=4} $. In summary, our IGIT-Net reconstructs the precise k-space spectrum and MRI images progressively by optimizing different loss $L_{stage_{j}}$ at various stages as follows:
\vspace{-0.1cm}
\begin{equation}
L_{stage_{j}}= {\textstyle \sum_{i=1}^{i=j}L_{i}} 
\label{eq20}
\end{equation}

\section{Experiments}
\subsection{Implementation Details}
\subsubsection{Datasets}
We use two real single-coil k-space datasets, CC359~\citep{CC359} and fastMRI~\citep{r28}, one simulated k-space dataset, IXI~\citep{IXI}, and one real multi-coil k-space dataset, SKM-TEA~\citep{SKMTEA}, to validate the effectiveness of our method.

\textbf{CC359 dataset} The brain MR raw dataset—the Calgary-Campinas dataset, which comes from a clinical MR scanner (Discovery MR750; GE Healthcare, Waukesha, WI, USA)—is used for validation. $4129$ slices from $25$ volumes are selected to form the training set, and the validation set is composed of $1650$ slices from $10$ other volumes.

\textbf{fastMRI dataset} The fastMRI dataset is currently the largest open-source MRI dataset with raw fully sampled k-space data. The single-coil knee dataset which consists of $973$ training volumes ($29877$ 2D slices) and $199$ validation volumes ($6140$ 2D slices) is employed in the experiments.

\textbf{IXI dataset} The IXI dataset consists of $574$ brain MRI volumes collected from three different hospitals in London. After pre-processing, $46226$ 2D slices from $368$ volumes are used for training, $11562$ 2D slices from $92$ volumes are used for validation, and $14315$ 2D slices from $114$ volumes are used for testing. Since IXI does not contain original k-space data, we obtain the simulated k-space via FFT.

\textbf{SKM-TEA dataset} The SKM-TEA raw data provides 155 multi-coil T2-weighted knee MRI scans, and each provides approximately 160 cross-sectional knee images with the matrix of size $512\times512$. Following the well-established ReconFormer ~\citep{ReconFormer}, we used 124, 10, and 21 volumes for training, validation, and testing, respectively.
\subsubsection{K-space undersampling}
In our experiments, two different types of masks are tested, i.e., 1D Cartesian mask and 2D Gaussian mask. For each mask, $20\%$ and $40\%$ undersampling rates are tested. Following \citep{r28}, all undersampling masks are generated by first including some number of adjacent lowest frequency k-space lines or points to provide a fully-sampled k-space region. Specifically, for the 1D Cartesian undersampling mask, the fully-sampled central region includes $8\%$ of all k-space lines; for the 2D Gaussian undersampling mask, $16\%$ of all k-space points are included. Then, the remaining k-space lines (for 1D Cartesian mask) or points (for 2D Gaussian mask) are included at random, with the probability set so that, on average, the undersampling mask achieves the desired acceleration factor.

\subsubsection{Baselines and Training Details}
We implement our model in PyTorch using Adam with an initial learning rate of $1e\-/3$. $(E_{0},E_{1},E_{2},E_{3},E_{4})$ for multi-stage training are set as $(0,20,60,100,200)$.

We compare our model with the following algorithms. Zero-Filling (ZF); ResNet\citep{r53}; UNet\citep{r28}; RefineGAN\citep{r54}, a generative adversarial model; SwinMR\citep{r34}, a swin Transformer based method; SwinGAN\citep{r38}, which combines the swin transformer and generative adversarial network for fast MRI; ReconFormer\citep{ReconFormer}, which designs a locally pyramidal but globally columnar transformer for multi-scale modeling; KIKI-Net\citep{r35}, which combines 4 CNNs to operate on k-space and image sequentially; D5C5\citep{r17D5C5}, a deep cascaded network for fast MRI. For fair comparison, we retrain these models using their default parameter settings. 

Peak Signal-to-Noise Ratio (PSNR), Structural Similarity Index Measure (SSIM), Normalized Mean Squared Error (NMSE), and Learned Perceptual Image Patch Similarity (LPIPS) are utilized for quantitative evaluation. 

\subsection{Experimental Results}
\subsubsection{Single-coil datasets}
The quantitative results of comparisons on single-coil datasets including CC359, fastMRI, and IXI datasets are shown in Table \ref{table.1}, Table \ref{table.2}, and Table \ref{table.3}, respectively. The best result in each column is highlighted in bold. Based on these results, we can find that: (1) Accelerated sampling leads to zero-filled MRI images with extremely poor quality; (2) Classical deep learning methods, such as UNet and ResNet, reduce image artifacts to some extent, but the restoration results remain suboptimal; (3) Current state-of-the-art methods, such as RefineGAN, SwinMR, and ReconFormer, have significantly improved image reconstruction quality, but their performance degrades on challenging tasks with high acceleration rates (sampling rate = 20\%). (4) Our method consistently achieves the best performance with significant improvements on all datasets under different sampling scenarios. For example, as shown in Table \ref{table.1}, our method shows the superiority of $1.08$ dB and $1.69$ dB in PSNR over ReconFormer under 20\% and 40\% 2D Gaussian mas sampling masks on the CC359 dataset, respectively. As shown in Table \ref{table.3}, our model outperforms ReconFormer by $0.51$ dB and $0.29$ dB in PSNR under 20\% and 40\% 1D Cartesia masks on the IXI dataset, respectively.

\subsubsection{Multi-coil dataset}
We further validate the effectiveness of our method on the multi-coil dataset SKM-TEA. The results in Table \ref{table4} show that our method still exhibits significant superiority over other competitors on the real multi-coil k-space dataset.

\subsubsection{Experiments on masks}
This experimental study aims to evaluate the robustness of our IGKR-Net using different undersampling trajectories. Five 1D Cartesian undersampling trajectories including 1D Cartesian $10\%$, 1D Cartesian $20\%$, 1D Cartesian $30\%$, 1D Cartesian $40\%$ , and 1D Cartesian $50\%$, as well as five 2D Random undersampling trajectories including 2D Random $10\%$, 2D Random $20\%$, 2D Random $30\%$, 2D Random $40\%$, and 2D Random $50\%$ are applied in this experiment. This experiment is conducted using the CC359 dataset.

\begin{figure*}[!t]
\centerline{\includegraphics[width=1.0\linewidth]{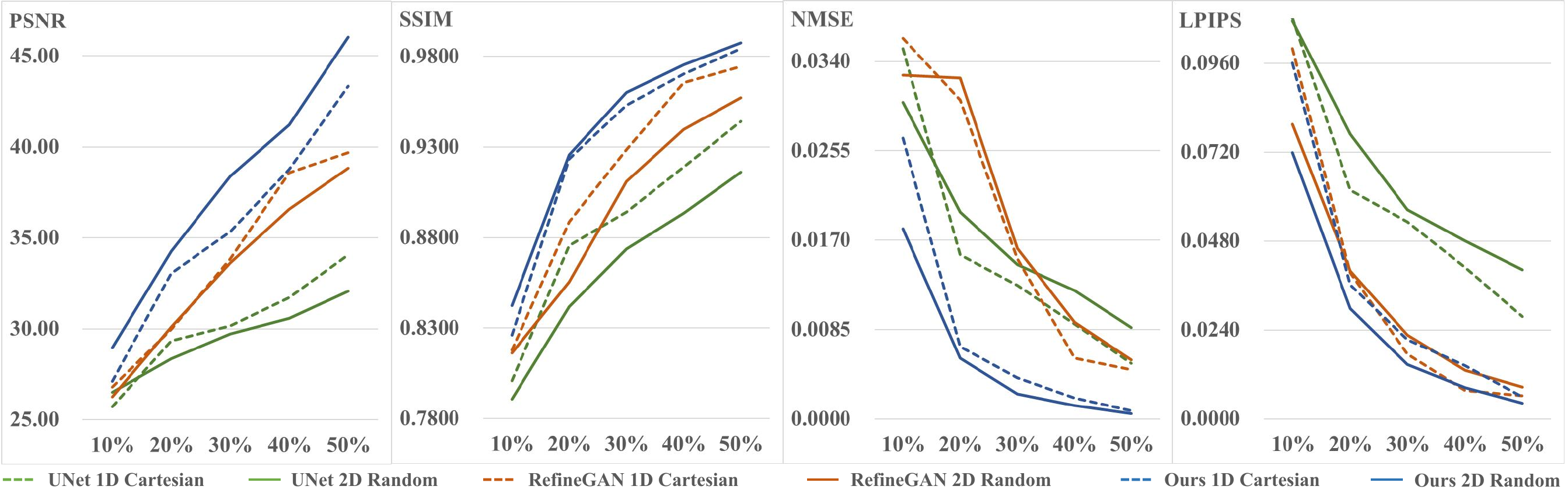}}
\caption{Results of the mask experiments. The X-axis denotes the sampling ratio, while the Y-axis corresponds to various metrics.}
\label{Fig.6}
\vspace{-0.5 cm}
\end{figure*}

The quantitative results are shown in Table \ref{table.5}. According to the results, we can fine that: (1) When using the same type of mask, e.g., 1D Cartesian mask, our IGKR-Net consistently demonstrates significant improvements across different sampling rates compared to other methods. For example, our model outperforms refineGAN by $3.66$ dB and $0.21$ dB under 1D Cartesian mask at $50\%$ and $40\%$ sampling rates, respectively. (2) As shown in Figure.\ref{Fig.6}, our method's performance under 2D masks (denoted as solid lines) even consistently surpasses that of other methods under 1D masks (denoted as dashed lines ) across different sampling rates, despiting 1D mask being recognized as a simpler task. (3) RefineGAN achieves satisfactory results at higher sampling rates, while its performance significantly declines under more challenging lower rates. Compared to RefineGAN, our method shows robustness across various masks.
\begin{table}[]
\renewcommand{\arraystretch}{1.1}
    \centering
    \small
    \setlength{\tabcolsep}{0.2mm}
    \caption{Performance comparison of MRI reconstruction under different undersampling masks on the multi-coil SKM-TEA.}
\begin{tabular}{cl|ccc|ccc}
\toprule
\multicolumn{2}{l|}{Mask type}                                                      & \multicolumn{3}{c|}{1D Cartesian}                                                                                          & \multicolumn{3}{c}{2D Random}                                                                                              \\ \midrule
\multicolumn{1}{c|}{ratio}                  & Method                                & PSNR↑                                   & SSIM↑                                    & NMSE↓                                    & PSNR↑                                   & SSIM↑                                    & NMSE↓                                    \\ \midrule
\multicolumn{1}{c|}{}                       & Unet                                  & 25.72                                  & 0.8010                                  & 0.0352                                  & 26.50                                  & 0.7909                                  & 0.0301                                  \\
\multicolumn{1}{c|}{}                       & RefineGAN                             & 20.80                                  & 0.5677                                  & 0.3890                                  & 26.27                                  & 0.7363                                  & 0.0327                                  \\
\multicolumn{1}{c|}{\multirow{-3}{*}{10\%}} & \cellcolor[HTML]{EFEFEF}\textbf{Ours} & \cellcolor[HTML]{EFEFEF}\textbf{27.11} & \cellcolor[HTML]{EFEFEF}\textbf{0.8262} & \cellcolor[HTML]{EFEFEF}\textbf{0.0267} & \cellcolor[HTML]{EFEFEF}\textbf{28.96} & \cellcolor[HTML]{EFEFEF}\textbf{0.8424} & \cellcolor[HTML]{EFEFEF}\textbf{0.0181} \\ \midrule
\multicolumn{1}{c|}{}                       & Unet                                  & 29.34                                  & 0.8756                                  & 0.0156                                  & 28.34                                  & 0.8417                                  & 0.0197                                  \\
\multicolumn{1}{c|}{}                       & RefineGAN                             & 29.96                                  & 0.8888                                  & 0.0303                                  & 30.10                                  & 0.8554                                  & 0.0324                                  \\
\multicolumn{1}{c|}{\multirow{-3}{*}{20\%}} & \cellcolor[HTML]{EFEFEF}\textbf{Ours} & \cellcolor[HTML]{EFEFEF}\textbf{33.06} & \cellcolor[HTML]{EFEFEF}\textbf{0.9234} & \cellcolor[HTML]{EFEFEF}\textbf{0.0069} & \cellcolor[HTML]{EFEFEF}\textbf{34.26} & \cellcolor[HTML]{EFEFEF}\textbf{0.9258} & \cellcolor[HTML]{EFEFEF}\textbf{0.0058} \\ \midrule
\multicolumn{1}{c|}{}                       & UNet                                  & 30.15                                  & 0.8940                                  & 0.0127                                  & 29.73                                  & 0.8738                                  & 0.0147                                  \\
\multicolumn{1}{c|}{}                       & RefineGAN                             & 33.84                                  & 0.9286                                  & 0.0152                                  & 33.62                                  & 0.9110                                  & 0.0162                                  \\
\multicolumn{1}{c|}{\multirow{-3}{*}{30\%}} & \cellcolor[HTML]{EFEFEF}\textbf{Ours} & \cellcolor[HTML]{EFEFEF}\textbf{35.35} & \cellcolor[HTML]{EFEFEF}\textbf{0.9530} & \cellcolor[HTML]{EFEFEF}\textbf{0.0039} & \cellcolor[HTML]{EFEFEF}\textbf{38.38} & \cellcolor[HTML]{EFEFEF}\textbf{0.9600} & \cellcolor[HTML]{EFEFEF}\textbf{0.0024} \\ \midrule
\multicolumn{1}{c|}{}                       & UNet                                  & 31.73                                  & 0.9189                                  & 0.0090                                  & 30.59                                  & 0.8935                                  & 0.0122                                  \\
\multicolumn{1}{c|}{}                       & RefineGAN                             & 38.57                                  & 0.9656                                  & 0.0058                                  & 36.58                                  & 0.9396                                  & 0.0092                                  \\
\multicolumn{1}{c|}{\multirow{-3}{*}{40\%}} & \cellcolor[HTML]{EFEFEF}\textbf{Ours} & \cellcolor[HTML]{EFEFEF}\textbf{38.78} & \cellcolor[HTML]{EFEFEF}\textbf{0.9703} & \cellcolor[HTML]{EFEFEF}\textbf{0.0020} & \cellcolor[HTML]{EFEFEF}\textbf{41.22} & \cellcolor[HTML]{EFEFEF}\textbf{0.9753} & \cellcolor[HTML]{EFEFEF}\textbf{0.0013} \\ \midrule
\multicolumn{1}{c|}{}                       & Unet                                  & 34.08                                  & 0.9442                                  & 0.0053                                  & 32.06                                  & 0.9160                                  & 0.0087                                  \\
\multicolumn{1}{c|}{}                       & RefineGAN                             & 39.68                                  & 0.9746                                  & 0.0047                                  & 38.84                                  & 0.9571                                  & 0.0056                                  \\
\multicolumn{1}{c|}{\multirow{-3}{*}{50\%}} & \cellcolor[HTML]{EFEFEF}\textbf{Ours} & \cellcolor[HTML]{EFEFEF}\textbf{43.34} & \cellcolor[HTML]{EFEFEF}\textbf{0.9844} & \cellcolor[HTML]{EFEFEF}\textbf{0.0008} & \cellcolor[HTML]{EFEFEF}\textbf{46.04} & \cellcolor[HTML]{EFEFEF}\textbf{0.9875} & \cellcolor[HTML]{EFEFEF}\textbf{0.0005} \\ \bottomrule
\end{tabular}
    \label{table.5}
\vspace{-1.5 em}
\end{table}
\subsubsection{Visualization Results}
The qualitative comparison results are shown in Figure.\ref{Fig.5}, including the reconstructed MRI images and absolute error maps. As observed, reconstructed MRI image using zero-filling (ZF) at high acceleration rates leads to significant aliasing artifacts and loss of anatomical details. In comparison to ZF, deep learning methods based on CNNs such as ResNet and UNet show improvement in reconstruction but still suffer from severe edge blurring and substantial detail loss. SwinGAN and ReconFormer partially alleviate these issues, yet in challenging tasks with high acceleration rates, these methods lose some crucial anatomical details. In contrast, our IGKR-Net demonstrates robustness to sampling patterns and acceleration rates. Leveraging information from the image domain to guide the recovery in k-space, our method better preserves important anatomical details, as indicated by the enlarged boxes and yellow ellipses.

\begin{figure}[!t]
\centerline{\includegraphics[width=1.0\linewidth]{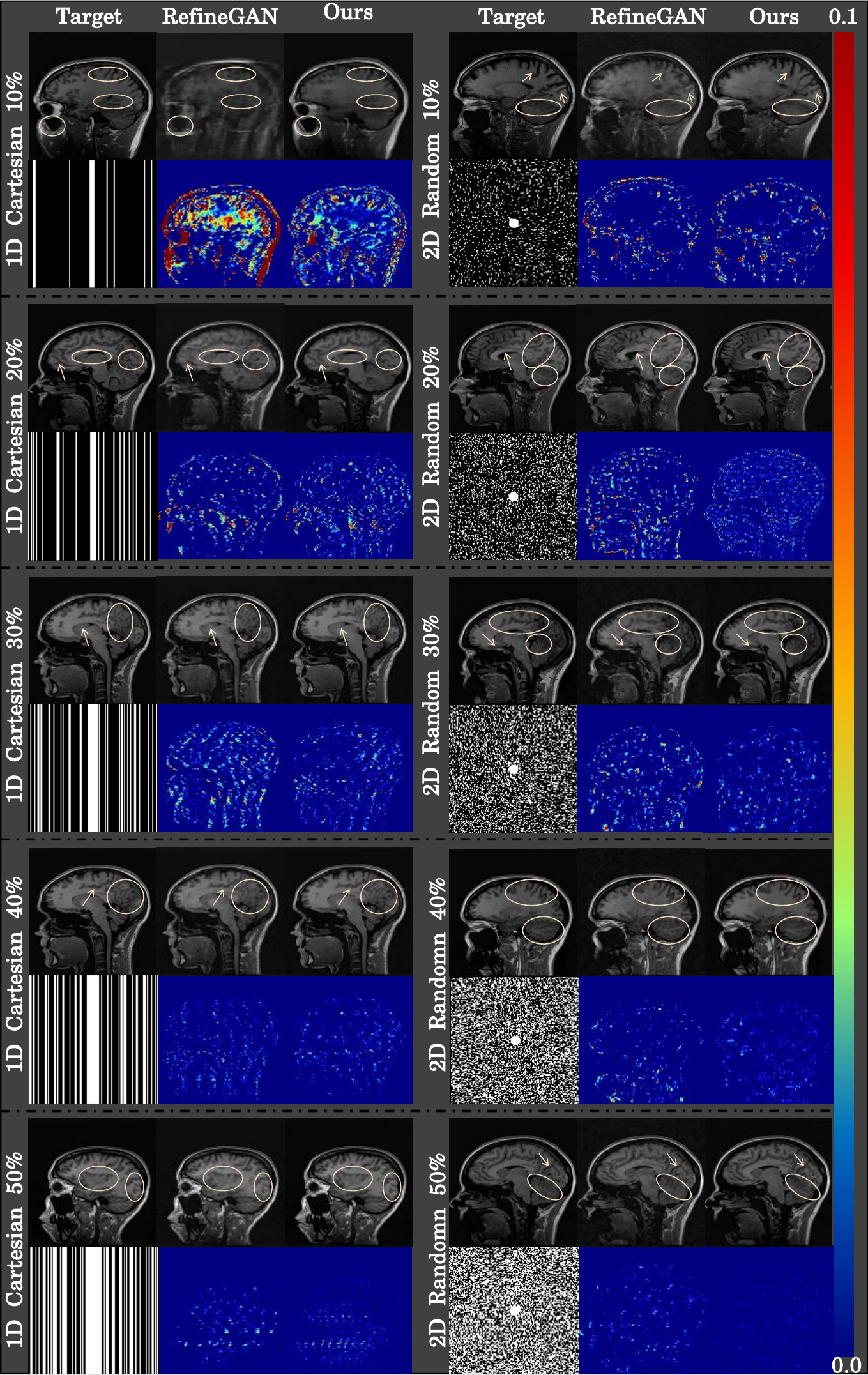}}
\caption{Visual comparison results of the mask experiments.}
\label{Fig.7}
\vspace{-0.5 cm}
\end{figure}

As shown in Figure.\ref{Fig.7}, as the k-space sampling ratio increases, the performance of different methods steadily improves. It is noteworthy that in challenging tasks with low sampling rates, e.g., 1D or 2D $10\%$, other methods obtain extremely poor performance, while our method can still reconstruct relatively clear contour edges. Overall, our IGKR-Net achieved a higher reconstruction quality compared to other methods under various undersampling trajectories.

\begin{table}[]
\renewcommand{\arraystretch}{1.1}
    \centering
    \small
    \setlength{\tabcolsep}{0.2mm}
    \caption{Quantitative results of the ablation studies using 1D Cartesian mask (sampling ratio = 20\%)  on the CC359 dataset.}
\begin{tabular}{l|cccc|cccc}
\toprule
Model & LRIT & HRIT & IDGM & TARM & PSNR↑  & SSIM↑   & NMSE↓   & LPIPS↓  \\ \midrule
(a)   &      & \pmb{$\checkmark$}    & \pmb{$\checkmark$}    & \pmb{$\checkmark$}    & 31.65 & 0.8996 & 0.0098 & 0.0552 \\
(b)   & \pmb{$\checkmark$}    &      & \pmb{$\checkmark$}    & \pmb{$\checkmark$}    & 28.56 & 0.7362 & 0.0535 & 0.1027 \\
(c)   & \pmb{$\checkmark$}    & \pmb{$\checkmark$}    &      & \pmb{$\checkmark$}    & 30.42 & 0.8826 & 0.0120 & 0.0612 \\
(d)   & \pmb{$\checkmark$}    & \pmb{$\checkmark$}    & \pmb{$\checkmark$}    &      & 31.97 & 0.9091 & 0.0086 & 0.0440 \\
\cellcolor[HTML]{EFEFEF}\textbf{Ours}  & \cellcolor[HTML]{EFEFEF}\pmb{$\checkmark$}    & \cellcolor[HTML]{EFEFEF}\pmb{$\checkmark$}    & \cellcolor[HTML]{EFEFEF}\pmb{$\checkmark$}    & \cellcolor[HTML]{EFEFEF}\pmb{$\checkmark$}    & \cellcolor[HTML]{EFEFEF}\textbf{33.06} & \cellcolor[HTML]{EFEFEF}\textbf{0.9234} & \cellcolor[HTML]{EFEFEF}\textbf{0.0069} & \cellcolor[HTML]{EFEFEF}\textbf{0.0362} \\ \bottomrule
\end{tabular}
    \label{table.6}
\vspace{-1.5 em}
\end{table}

\subsection{Ablation Studies and Further Analysis}
\subsubsection{Efficacy of Key Components}
To validate the effectiveness of the Key Components within our proposed method, we conduct a breakdown ablation experiment, the results are shown in Table \ref{table.6}. We can observe that: (a) When removing the low- to high-resolution progressive learning strategy, i.e., querying the unsampled k-values directly using high-resolution coordinates, the model produces a significant performance degradation from PSNR $33.06$ to $31.65$. Direct recovery of high-resolution k-space from overly undersampled data leads to oversmoothing, which results in poorer reconstruction results. This illustrates the effectiveness of our low- to high-resolution multi-stage learning strategy. (b) Disabling the HRIT module, i.e., directly up-sampling the low-resolution outputs obtained by LRIT as the reconstruction results, resulted in a severe decrease of (PSNR, SSIM) by $(4.50, 0. 1872)$, respectively. This suggests that using the dense k-space obtained from LRIT to further construct a more accurate latent space, and querying the k-space by high-resolution coordinates is important for effective MRI reconstruction.
(c) When removing IDGM from the network and using the k-space spectrum outputted from the LRIT for HRIT, the performance decreased significantly, which demonstrates that our proposed IDGM can mine information from undersampled MRI images to obtain more informative outputs. (d) Disabling the TARM led to a $1.09$ drop in PSNR and $0.0143$ decrease in SSIM, demonstrating its importance in refining the reconstruction outputs in image domain, thereby achieving accurate results.
\vspace{-1.0em}
\subsubsection{Efficacy of Recovering K-space}
\begin{figure}[!t]
\centerline{\includegraphics[width=1.0\linewidth]{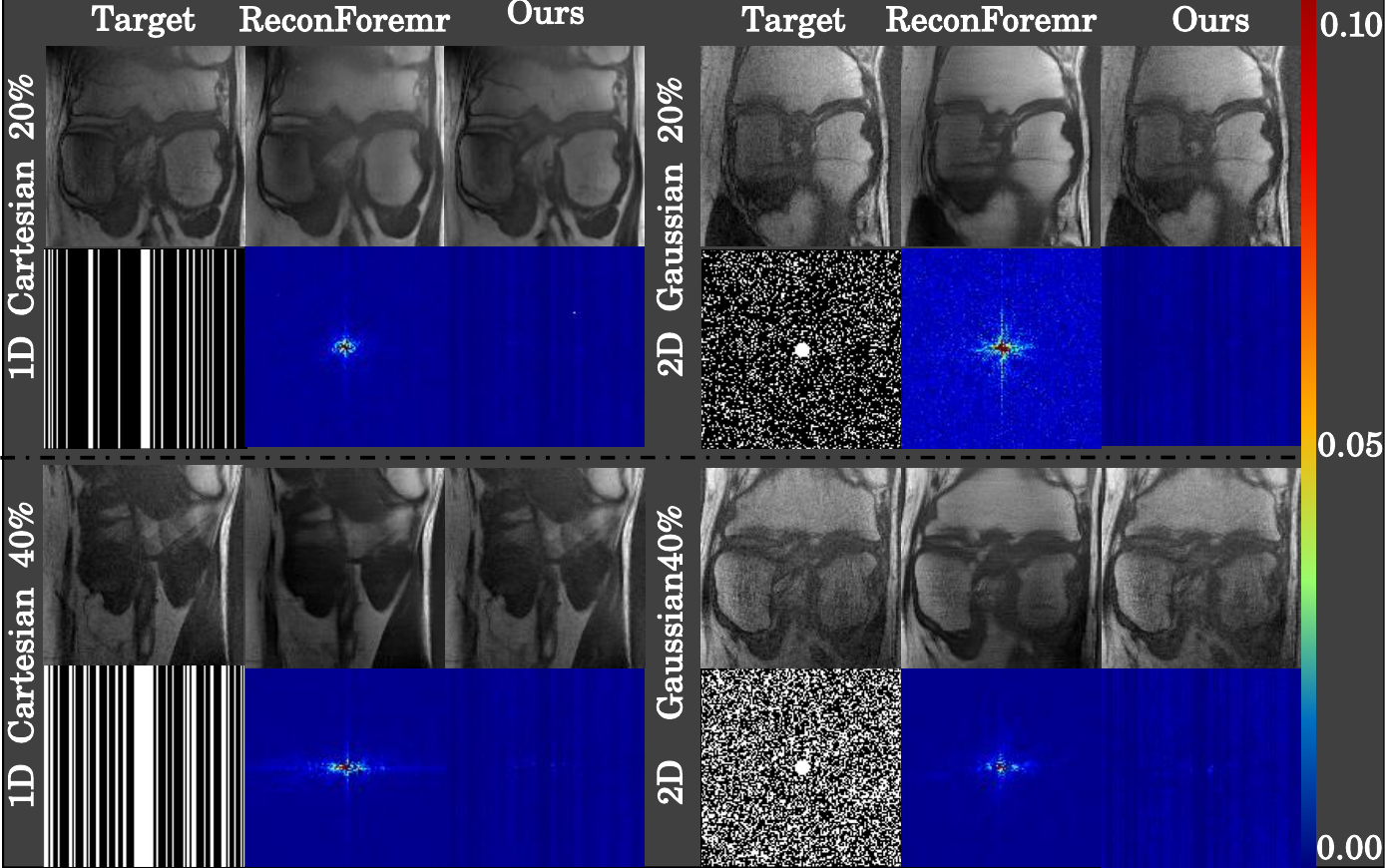}}
\caption{Visual comparison results with k-space error maps.}
\label{Fig.8}
\vspace{-1.5em}
\end{figure}

To further analyze the effectiveness of our approach in the recovery of k-space, we present the results of k-space reconstruction, as shown in Figure. \ref{Fig.8}. Specifically, we show the sample of reconstructed images and absolute differences ($10\times$) of standardized pixel intensities between reconstructed k-space and GT ones. It can be observed that our method achieves a generally low error in the recovered spectrum. Corresponding to the MRI images, it is evident that our approach excels in recovering high-frequency details and edges, while other methods tend to exhibit issues of over-smoothing. This further substantiates the importance of k-space recovery for the precise reconstruction of MRI images and the effectiveness of our proposed method.

\begin{table}[!h]
\renewcommand{\arraystretch}{1.1}
\centering
\setlength{\tabcolsep}{0.7mm}
    \caption{Efficiency comparison on CC359 dataset. (1D Cartesian 20\%)}
    \scalebox{0.94}
    {
\begin{tabular}{l|cc|cccc}
\toprule
Method        & FLOPs & Param. & PSNR↑  & SSIM↑   & NMSE↓   & LPIPS↓           \\ \midrule
SwinMR        & 398G  & 11.40M & 31.11          & 0.9006          & 0.0124          & 0.0413          \\
ReconFormer   & 342G  & 1.14M  & 32.79          & 0.9189          & 0.0072          & 0.0373          \\
\rowcolor[HTML]{EFEFEF} 
\textbf{Ours} & 138G  & 13.89M & \textbf{33.06} & \textbf{0.9234} & \textbf{0.0069} & \textbf{0.0362} \\ \bottomrule
\end{tabular}
    }
    \label{table.7}
   \vspace{-1.0em}
\end{table}
\subsubsection{Analysis of Training Efficiency}
The results of the training efficiency comparison are reported in Table \ref{table.7}. SwinMR obtained a small number of parameters but with a large computational complexity (FLOPs=$398$G) and suboptimal reconstruction results. ReconFormer used a local pyramidal but global columnar transformer structure, which achieved a lightweight parameter of $1.14$M via its recurrent nature. However, the larger computational complexity (FLOPs = $342$G) hinders its training and testing process. Our IGKR-Net outperformed other methods and obtained an optimal balance between the number of parameters and computational complexity.
\section{Discussion and Future Work}
This paper introduces IGKR-Net, a network that addresses the challenging problem of k-space recovery caused by under-sampling, using INR as a new perspective. Furthermore, by introducing an image guidance module that extracts information from low-quality MRI images and proposing a tri-attention refinement module that enhances the resulting image quality, IGKR-Net ultimately achieves accurate MRI reconstruction. Experimental results demonstrate our method's effectiveness, showing superior performance compared to existing MRI reconstruction techniques on various datasets. In future work, we will also extend IGKR-Net to other recovery tasks, such as cardiac MRI.
\section{Conclusion}
In this paper, we focused on the problem of image-domain guided k-space recovery for MRI reconstruction. We proposed a novel method, called IGKR-Net, which tackled the key problem of current deep learning based MRI reconstruction methods, namely the insufficient and unsuitable k-space recovery. Specifically, an implicit transformer based k-space reconstruction is proposed to efficiently learn continuous feature space and recover high-quality k-space in the Fourier domain. This mechanism facilitates more precise and effective learning of continuous representations of k-space information, resulting in precise k-space spectrum that are more favorable for MRI reconstruction. Furthermore, an image domain enhancement branch consists of a fusion module and a refinement module is introduced to enable accurate dual-domain feature aggregation and image refinement, thereby guiding MRI reconstruction. Our method demonstrates superior performance on publicly available datasets, confirming its effectiveness.

\bibliographystyle{cas-model2-names}
\bibliography{reference}

\end{document}